\documentclass[twocolumn,showpacs,showkeys,preprintnumbers,amsmath,amssymb]{revtex4}

\usepackage{graphicx}

\begin{document}

\title{Geometry dependence of the charge transfer at YBa$_2$Cu$_3$O$_7$-metal interfaces}

\author{C.~Schuster}
\author{U.~Schwingenschl\"ogl}
\affiliation{Institut f\"ur Physik, Universit\"at Augsburg, 86135 Augsburg, Germany}

\date{\today}

\begin{abstract}
Transport in electronic devices based on high-T$_c$ superconductors
depends critically on the charge redistribution at interfaces, since
the band structure is modified on a local scale. Using the density
functional theory approach for relaxed YBa$_2$Cu$_3$O$_7$-metal
contacts, the charge transfer into the superconductor has been
studied in Appl.\ Phys.\ Lett.\ {\bf 90}, 192502 (2007). In the
present paper we discuss the systematics inherent in the local electronic
structure  of the near-contact YBa$_2$Cu$_3$O$_7$ sites, in
particular the dependence on the contact geometry.
\end{abstract}



\maketitle

\section{Introduction}

Due to large dielectric constants and small carrier \cite{samara90}
densities, electronic transport in wires and tapes from high-T$_c$
materials is seriously affected by structural defects and interfaces
\cite{mannhart98,hilgenkamp02}. Moreover, the specific contact
resistivity of YBa$_2$Cu$_3$O$_7$(YBCO)-metal thin films is determined by details of the
contact geometry \cite{hahn94} and the transport in micron-sized
YBCO-metal heterojunctions depends on the orientation of the YBCO
crystallographic axes with respect to the direction of the current
flow \cite{komissinskii01}. Unexpected high specific resistivities
of YBCO-Au interfaces, $10^{-4}\,\Omega\,{\rm cm}^2$ to
$10^{-3}\,\Omega\,{\rm cm}^2$ at low temperatures \cite{xu04},
presumably result from local distortions at the interface. While
theoretical attempts to describe the charge distribution at such
interfaces are rare, the linear background and asymmetry in the
differential conductance of a tunnel junction between a high-T$_c$
superconductor and a metal can be explained by inelastic scattering
\cite{grajar97}. In addition, the charge imbalance at the boundary
of a short coherence length superconductor and a normal metal is
accessible to a self-consistent microscopic approach \cite{nikolic02}.

\section{Technical details}
Taking into account the details of the crystal structure, the
electronic properties of YBCO-metal interfaces can be
addressed by means of first principles band structure calculations
\cite{us07c}. In this paper we investigate the dependence of the
local electronic structure in the YBCO domain on the contact geometry.
As YBa$_2$Cu$_3$O$_7$ belongs to the overdoped regime, our interfaces
are metallic. The calculations are based on density functional theory
and the generalized gradient approximation, as implemented in the
WIEN$2k$ program package \cite{wien2k}, which is suitable for dealing
with structural relaxation and charge
redistribution at surfaces and interfaces \cite{us07a,us07b}. Because
band-bending is proposed to take place on the length scale of the
YBCO lattice constant, the electronic structure of YBCO-metal interfaces
becomes accessible to a supercell approach with periodic boundary
conditions. Due to a minimal lattice mismatch of $\approx0.7$\%, it
is convenient to choose fcc Pd as metallic substituent.
\begin{figure}
\includegraphics[width=0.45\textwidth,clip]{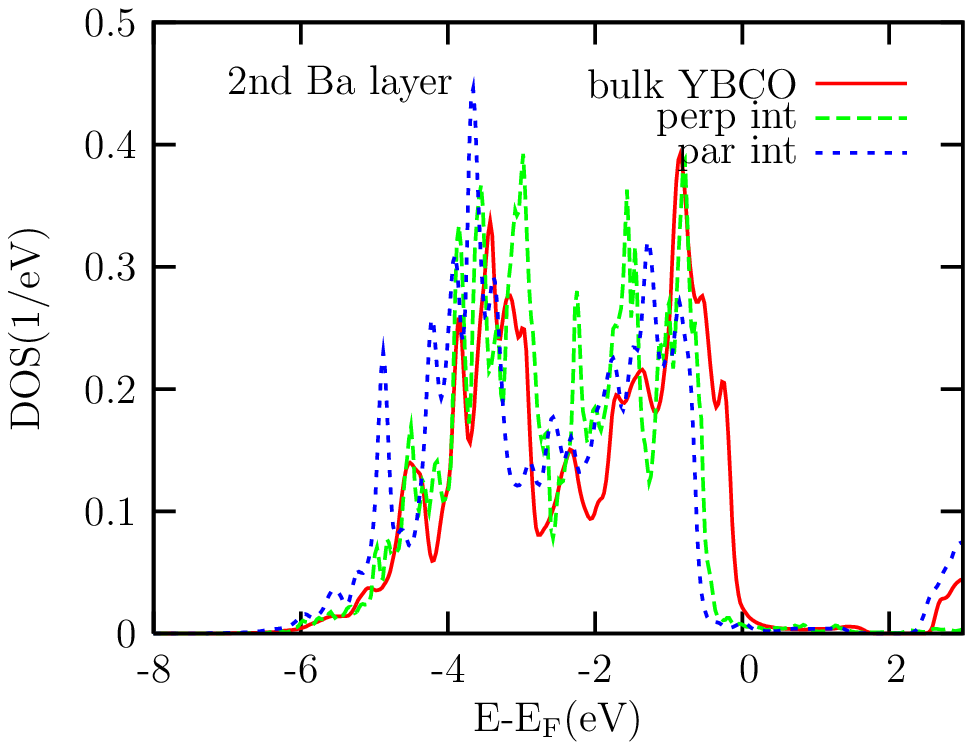}\\
\includegraphics[width=0.45\textwidth,clip]{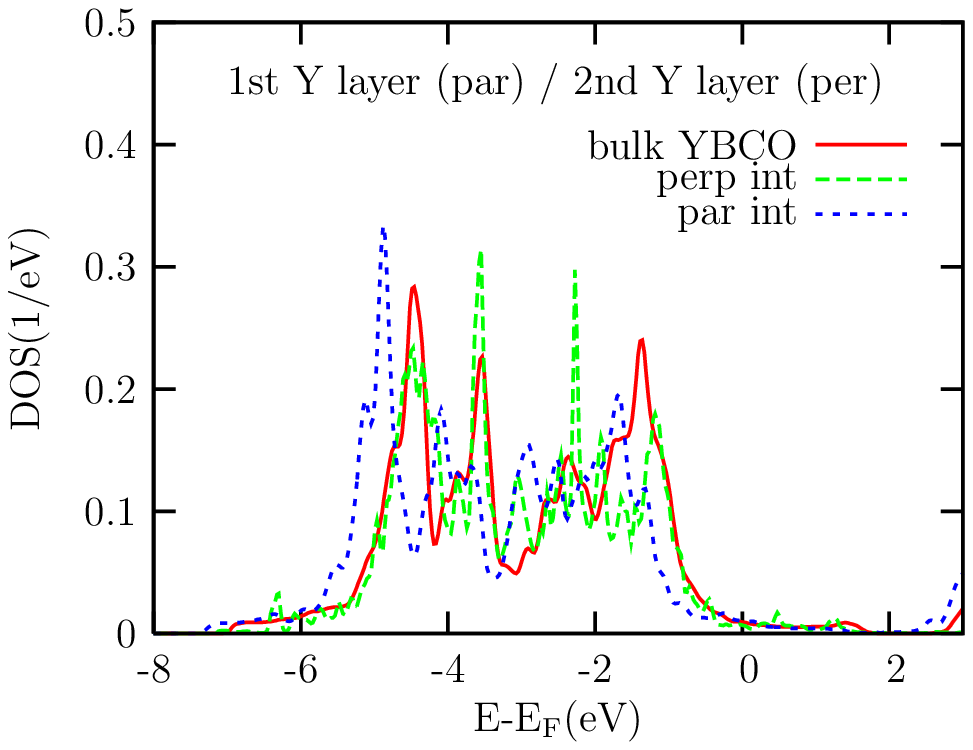}
\caption{Partial Ba and Y densities of states.}
\label{fig-Ba}
\end{figure}

We study interfaces oriented parallel and perpendicular to the
CuO$_2$ planes. The parallel interface is modelled by means of a
supercell consisting of two YBCO unit cells along the crystallographical
$c$-axis, terminated by the CuO-chain layer \cite{xin89,derro02}, and
four metallic unit cells stacked in [001] direction. However, the
results are closely related to those for BaO terminated YBCO. The
perpendicular interface consists of three YBCO unit cells and a
$3\times3$ metallic supercell stacked in [010] direction. In total,
the supercell therefore contains 9 metallic unit cells. We again
choose a CuO/Pd interface to prepare for comparison with the parallel
configuration. In contrast to the latter, oxygen atoms from the BaO
planes and copper atoms from the CuO$_2$ planes here likewise touch
the metal layer. Giving rise to the main difference between the two
configurations, in the parallel configuration the CuO$_2$ planes are
well separated from the metal by an almost insulating BaO layer,
whereas they are in contact in the perpendicular configuration. The Y
atoms are in touch with the metal layer only in the latter case.
Relative shifts between the YBCO and metal domain parallel to the
interface have minor effects. We start from the experimental bulk
YBCO lattice constants $a=3.865$\,\AA\ and $b=3.879$\,\AA\ \cite{siegrist87},
and optimize the atomic coordinates in a first step \cite{kouba99}.
Parallel to the interface, the YBCO lattice constants are used for
the palladium domain, too. In the perpendicular direction, we apply
the fcc Pd lattice constant $c=3.89 $\,\AA. In a second step, the
structural relaxation of the supercells is carried out, see \cite{us07d}
for details of the structure optimization.

\section{Results}

In the following discussion, we compare the electronic structures
of our YBCO-metal interfaces with results of a bulk YBCO calculation
(which agree perfectly with previous findings \cite{pickett89}).
In addition, we will concentrate on atomic sites slightly off the
YBCO-metal contact, in order to determine a common behaviour. These
sites are less affected by the structural optimization, whereas the
electronic properties of atoms right at the interface depend strongly
on structural subtleties and therefore are not of interest here. We
first study the partial density of states (DOS) associated with the
Ba and Y electron donor sites, see the DOS curves in fig.\ \ref{fig-Ba}.
For both geometries, the 2nd Ba layer off the interface is well
separated from the metal domain. As a consequence, the gross features
of the DOS curves agree well with each other as well as with the bulk
YBCO DOS, when an adequate energetical shift of about $-0.45$\,eV is
applied. In contrast, the Y DOS curves hardly coincide, which traces
back to large differences in the local atomic environments of the Y
sites in the parallel and perpendicular interface configuration. The
1st Y layer off the parallel interface is well separated from the
metal by four atomic layers. The energetical shift of the Y states
therefore can be attributed to strong down-bending of the electronic
bands due to a modified Fermi level. At the perpendicular interface,
Y-Pd orbital overlap suppresses band-bending effects. While the Y DOS
reveals a finite number of states at the Fermi level, the Ba DOS
almost vanishes. Electronic screening thus is less efficient at the
Ba sites and the band-bending amplitude increases. 
\begin{figure}
\includegraphics[width=0.45\textwidth,clip]{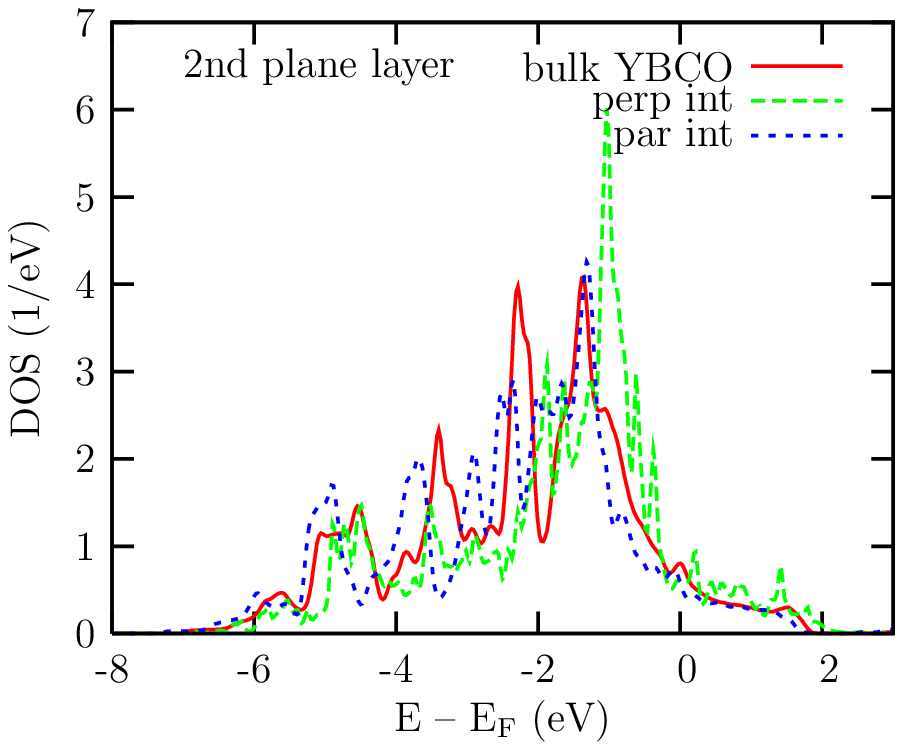}\\
\includegraphics[width=0.45\textwidth,clip]{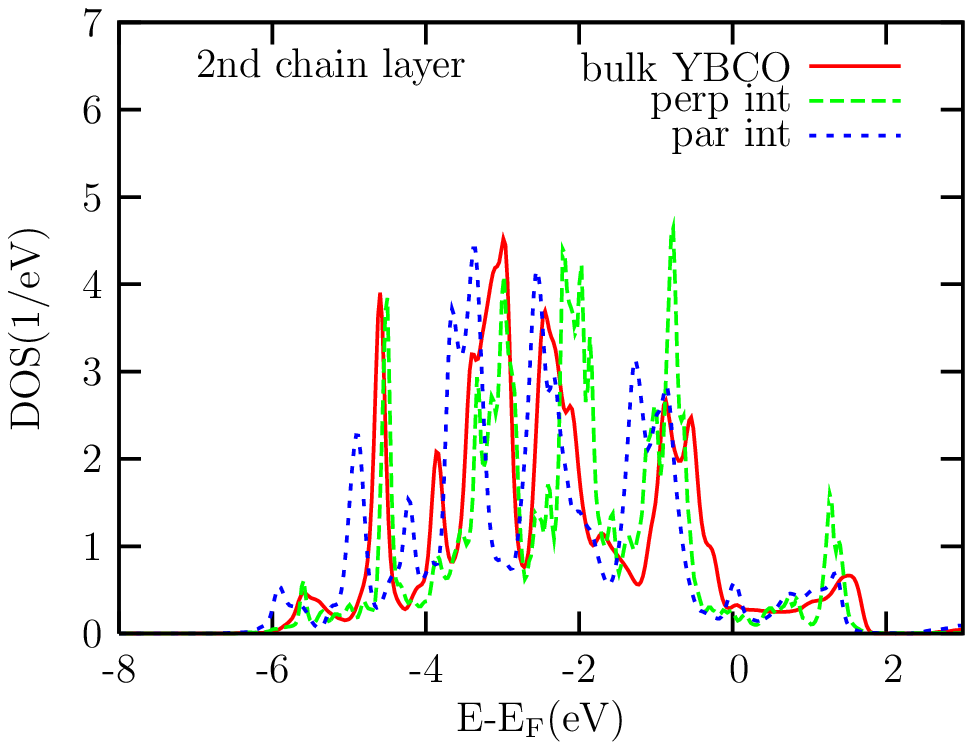}
\caption{Partial Cu $3d$ densities of states for the CuO$_2$-planes
and CuO-chains.}
\label{fig-Cu}
\end{figure}

Turning to the copper sites, the gross structure of the Cu $3d$ DOS
is hardly affected by the interface, see fig.\ \ref{fig-Cu}. The
second layer Cu atoms largely resemble the bulk DOS and therefore
allow us to study the band-bending. Concerning the CuO-chain sites,
results for the parallel and perpendicular interface configuration
agree quite well. Similar to the discussion of the Y DOS, the smaller
energy shift in the perpendicular case can be explained by direct
Cu-Pd orbital overlap. Understanding the DOS shape of the CuO$_2$-plane
sites is less intuitive. For the parallel interface, we observe almost
perfect agreement of the DOS curves, whereas some effects of the
structural relaxation are left for the perpendicular interface.
However, in both cases the bulk DOS has to be shifted to lower energies
in order to reconcile the curves, which we again interpret in terms
of down-bending of the electronic bands. The necessary shifts amount
to 0.20\,eV and 0.15\,eV in the parallel and perpendicular case,
respectively. As a consequence, the hole count at the Cu sites in the
CuO$_2$-planes is altered, where the 0.20\,eV shift corresponds to a
reduction of 0.13 holes and the 0.15\,eV shift comes along with a
reduction of 0.09 holes. Since the charge transfer depends only
little on the orientation of the YBCO-metal interface with respect
to the high-T$_c$ unit cell, an intrinsic doping close to 0.1\,eV
seems to be characteristical for YBCO-metal contacts. As a consequence,
the system is shifted into the underdoped regime of the YBCO phase
diagram, which is expected to be reflected by transport properties,
like the current in tunneling experiments.

\section{Conclusion}

In conclusion, we have discussed electronic structure calculations
for prototypical interfaces between the short coherence
length superconductor YBa$_2$Cu$_3$O$_7$ and a normal metal.
We find that the charge redistribution
induced by the metal contact can be interpreted in terms of an
intrinsic doping of the superconductor on a nanometer length scale.
This fact corresponds well with the experimental observation of
charge carrier depletion. The net charge transfer in favour of the
copper sites in the CuO$_2$-planes amounts to some 0.1 electrons,
with only a weak dependence on the orientation of the interface. We
expect that the reported charge transfer amplitude applies to any
contact geometry as long as the YBCO domain terminates with a
copper/oxygen layer.

\begin{acknowledgments}
We acknowledge valuable discussions with U.\ Eckern, V.\ Eyert, J.\ Mannhart,
and T.\ Kopp, and financial support by the Deutsche Forschungsgemeinschaft
(SFB 484).
\end{acknowledgments}

\end{document}